\documentclass[aps,prd,reprint,superscriptaddress]{revtex4-2}
\usepackage{amsmath,amssymb,graphicx,bm,braket,physics}
\usepackage[colorlinks=true,linkcolor=blue,citecolor=blue,urlcolor=magenta]{hyperref}
\usepackage{color}
\usepackage{mathtools}
\usepackage{enumitem}
\usepackage{epsfig}
\usepackage{booktabs}
\providecommand{\Tr}{\mathrm{Tr}}
\newcommand{\inlinesection}[1]{\textit{#1}---}

\begin{document}
	\title{Multiparameter sensing of axion dark matter with superconducting-qubit networks}
	\author{Le Bin Ho}
	\email{ho.bin.le.e3@tohoku.ac.jp}
	\affiliation{Department of Applied Physics, Graduate School of Engineering, Tohoku University, Sendai 980-8579, Japan}
	\affiliation{Frontier Research Institute for Interdisciplinary Sciences, Tohoku University, Sendai 980-8578, Japan}
	\date{\today}

\begin{abstract}
We propose a quantum sensor network for direct detection of quantum chromodynamics (QCD) axion dark matter using entangled superconducting qubits. In the presence of a static magnetic field, the oscillating axion field induces a coherent qubit rotation described by an encoding angle proportional to the axion-photon coupling and an unknown dark-matter phase. Treating the phase as an additional unknown parameter turns axion detection into a two-parameter quantum estimation problem. We formulate these signals in Cartesian coordinates, thereby avoiding the singularity that arises in the weak-signal regime and establishing a regular framework for multiparameter quantum estimation. Using this framework, we optimize the quantum sensor network and combine it with Bayesian inference to reconstruct the axion-photon coupling. The optimized protocol accurately recovers the KSVZ and DFSZ benchmark couplings over the mass range $m_a\in[0.1,10]~\mu{\rm eV}$, generalizes to previously unseen masses, and remains robust against realistic superconducting-qubit decoherence and gate errors.
\end{abstract}

\maketitle

\inlinesection{Introduction}
The quantum chromodynamics (QCD) axion, originally proposed to solve the strong charge-parity (strong-CP) problem~\cite{PhysRevLett.38.1440,PhysRevLett.40.223,PhysRevLett.40.279}, remains one of the leading candidates for cold dark matter and continues to motivate both experimental searches and quantum simulation studies~\cite{yx1t-mkgp}. Its coupling to photons, $g_{a\gamma\gamma}$, has motivated a broad range of experimental searches, including resonant microwave haloscopes~\cite{PhysRevLett.120.151301,PhysRevLett.127.261803,PhysRevLett.126.191802,PhysRevD.107.072007} and microwave cavity detectors employing single-photon detection~\cite{PhysRevD.88.035020}. Recent advances in quantum technologies have further expanded this landscape, leading to proposals based on atomic clocks and quantum metrology~\cite{PhysRevA.111.012601}, multilevel quantum sensors~\cite{zvzb-yv67}, and error-corrected spin-qubit arrays~\cite{2y5c-x1kz}.

Among the available quantum platforms, superconducting transmon qubits have emerged as a promising candidate for axion detection. Originally proposed for hidden-photon dark matter~\cite{PhysRevLett.131.211001} and subsequently extended to QCD axions~\cite{PhysRevD.110.115021}, the scheme exploits the interaction between the axion-induced electromagnetic field and a transmon qubit to generate a weak coherent qubit rotation in the presence of a strong static magnetic field. Combined with cavity enhancement and Greenberger--Horne--Zeilinger (GHZ) entanglement~\cite{10.1119/1.16243,PhysRevLett.133.021801}, it can in principle probe the parameter space predicted by the Kim--Shifman--Vainshtein--Zakharov (KSVZ)~\cite{PhysRevLett.43.103,SHIFMAN1980493} and Dine--Fischler--Srednicki--Zhitnitsky (DFSZ)~\cite{DINE1981199,Zhitnitsky1980} axion models. Related transmon-based proposals have also been developed for direct dark-matter detection~\cite{PhysRevLett.126.141302,PhysRevX.15.021031,9p1t-vc9j}.
Recently, a variational quantum-sensor-network framework has been proposed for optimizing transmon-qubit networks through direct maximization of the quantum and classical Fisher information matrices (QFIM and CFIM)~\cite{rv43-54zq}.

Most existing protocols implicitly assume a known dark-matter phase. In practice, however, the axion field remains coherent only over a finite coherence time, after which the phase changes randomly. As a result, realistic axion detection must treat the phase as an additional unknown parameter, transforming the problem into one of multiparameter quantum estimation.

In this Letter, we develop the quantum sensor network in Ref.~\cite{rv43-54zq} for axion dark-matter detection in the presence of an unknown random dark-matter phase. Rather than treating the phase as the quantity of interest, we explicitly incorporate it as an additional unknown parameter when estimating the axion-photon coupling, thereby reformulating the detection task as a two-parameter quantum estimation problem. We show that describing the unknown signal in terms of its magnitude and phase causes the QFIM to become singular in the weak-signal regime. This singularity is not a physical limitation, but a consequence of the chosen coordinate representation. Reformulating the estimation problem in Cartesian coordinates removes the singularity while preserving the underlying physics, yielding a well-defined framework for multiparameter quantum estimation.

Within this framework, we optimize the network configuration and then employ Bayesian inference to reconstruct the axion-photon coupling from measurement outcomes. The proposed framework accurately recovers the KSVZ and DFSZ benchmark couplings over the full target mass range and generalizes to previously unseen axion masses. Under a realistic superconducting-qubit noise model, the quantum sensor network naturally favors shallower architectures with fewer entangling gates while preserving reliable estimation performance.

\begin{figure}[t]
\centering
\includegraphics[width=\columnwidth]{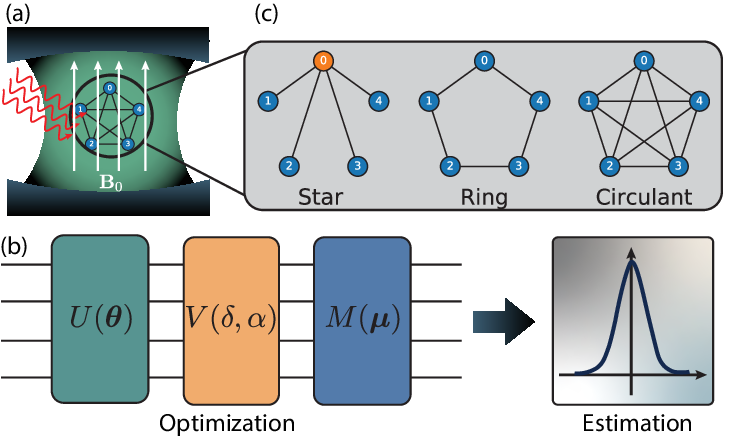}
\caption{
\textbf{Overview of the variational quantum sensor network for axion detection.}
(a) A static magnetic field $\mathbf{B}_0$ converts the oscillating axion dark-matter field into an effective electric field that drives an entangled network of transmon qubits.
(b) Variational sensing protocol: A parameterized circuit $U(\bm{\theta})$ prepares the entangled sensor state, followed by the axion-induced unitary $V(\delta,\alpha)$, a variational measurement circuit $M(\bm{\mu})$, and computational-basis readout for parameter estimation.
(c)  Sensor-network topologies considered in this work: star, ring, and circulant.
}
\label{fig:1}
\end{figure}

\inlinesection{Axion-transmon interaction model}
The QCD axion couples to electromagnetism through the interaction \cite{PhysRevLett.51.1415}
\begin{align}
\mathcal{L}\supset
g_{a\gamma\gamma}a\,\mathbf{E}\!\cdot\!\mathbf{B}.
\end{align}
In the presence of a static bias field $\mathbf{B}_0$, an oscillating axion field
$
a(t)=a_0\cos(m_at-\alpha),
$
with
$
a_0=\sqrt{2\rho_{\rm DM}}/m_a,
$
induces an effective oscillating electric field with amplitude
$
\bar E=g_{a\gamma\gamma}a_0B_0\kappa,
$
where $\rho_{\rm DM}$ is the local dark-matter density, $\alpha$ is an unknown phase, and $\kappa=O(1)$ is a cavity form factor~\cite{PhysRevD.110.115021} as illustrated in Fig.~\ref{fig:1}(a).

The induced electric field drives a capacitively coupled transmon qubit with coupling strength
\(
\eta=\frac{\sqrt{\omega_qC}}{2\sqrt{2}}\,
\bar Ed,
\)
where $C$ and $d$ are the transmon capacitance and electrode spacing, and $\omega_q$ is the qubit transition frequency. Near resonance ($m_a\simeq\omega_q$), the rotating-wave approximation yields
\begin{equation}
H_{\rm DM}
=
\hbar\eta
(\cos\alpha\,\sigma_x+\sin\alpha\,\sigma_y),
\label{eq:Hdm}
\end{equation}
corresponding to a resonant qubit rotation about an axis in the $xy$ plane determined by the axion phase $\alpha$. After an interrogation time
$\tau=\min(\tau_{\rm DM},T_2)$,
where
$\tau_{\rm DM}\sim1/(m_av_{\rm DM}^2)$
is the dark-matter coherence time and
$T_2$ the qubit coherence time, the evolution is
\begin{equation}
U_{\rm DM}(\delta,\alpha)=
\begin{pmatrix}
\cos\delta &
-ie^{-i\alpha}\sin\delta\\
-ie^{i\alpha}\sin\delta &
\cos\delta
\end{pmatrix},
\
\delta=\eta\tau,
\label{eq:UDM}
\end{equation}
where the rotation angle $\delta$ encodes the axion-photon coupling, while the unknown phase $\alpha$ determines the rotation axis. Since $\alpha$ remains constant only within one dark-matter coherence interval and is randomized between intervals, it must be treated as an additional unknown parameter in the estimation of $\delta$.

Throughout this work we adopt the benchmark parameters of Ref.~\cite{PhysRevD.110.115021}:
$B_0=5$~T,
$C=0.1$~pF,
$d=100~\mu$m,
$\kappa=1$,
$T_2=100~\mu$s,
$\rho_{\rm DM}=0.45~{\rm GeV/cm^3}$,
and
$v_{\rm DM}=10^{-3}c$.
For
$m_a\in[0.1,10]~\mu{\rm eV}$,
the coherence time exceeds $T_2$, so $\tau=T_2$ throughout and
\begin{equation}
\delta(g_{a\gamma\gamma},m_a)
\simeq
1.658
\left(
\frac{g_{a\gamma\gamma}}
{10^{-10}\ {\rm GeV}^{-1}}
\right)
\left(
\frac{m_a}
{1~\mu{\rm eV}}
\right)^{-1/2}.
\label{eq:delta2g}
\end{equation}
This formula provides the mapping between the measurable rotation angle $\delta$ and the axion-photon coupling. 

For concreteness we benchmark against the KSVZ~\cite{PhysRevLett.43.103,SHIFMAN1980493} and DFSZ~\cite{DINE1981199,Zhitnitsky1980} axion models, for which
\begin{equation}
g_{a\gamma\gamma}
=
\frac{\alpha_{\rm em}}{2\pi f_a}
\left|\frac{E}{N}-1.92\right|,
\label{eq:gagg}
\end{equation}
where $\alpha_{\rm em}\simeq1/137.036$ is the fine-structure constant, with $E/N=0$ for KSVZ and $E/N=8/3$ for DFSZ, and $f_a\simeq5.69\times10^{12}~{\rm GeV}\,(\mu{\rm eV}/m_a)$, so that both couplings scale linearly with $m_a$.

\inlinesection{Multiparameter estimation}
To estimate the axion parameters $(\delta,\alpha)$, we employ a variational quantum sensor network of $N$ transmon qubits~\cite{rv43-54zq}, as illustrated in Fig.~\ref{fig:1}(b). A variational circuit $U(\bm{\theta})$ first prepares an entangled sensor state, after which the axion encoding
$
V(\delta,\alpha)=U_{\rm DM}(\delta,\alpha)^{\otimes N}
$
is applied identically to all qubits. The encoded state is then processed by a variational measurement circuit $M(\bm{\mu})$, followed by computational-basis readout for parameter estimation.

Throughout this work, we consider a five-qubit sensor network ($N=5$) with three graph topologies shown in Fig.~\ref{fig:1}(c): star, ring, and circulant. In the star topology, a central qubit is connected to all others. In the ring topology, each qubit is connected to its two nearest neighbors. In the circulant topology, each qubit is connected to its first- and second-nearest neighbors, which is equivalent to a fully connected graph for $N=5$. The state-preparation and measurement circuits consist of $L_1$ and $L_2$ variational layers, respectively.

We quantify sensitivity with the quantum Fisher information matrix (QFIM) $\mathbf{Q}=[Q_{jk}]$, whose elements are
\begin{equation}
Q_{jk}
=
2
\sum_{n,m:\,\lambda_n+\lambda_m>0}
\frac{
\mathrm{Re}
\!\left[
\langle n|\partial_j\rho|m\rangle
\langle m|\partial_k\rho|n\rangle
\right]
}
{\lambda_n+\lambda_m},
\label{eq:QFIM}
\end{equation}
where $j,k\in\{\delta,\alpha\}$ and
$\rho=\sum_n \lambda_n|n\rangle\langle n|$
is the final quantum state of the sensing protocol.
The corresponding classical Fisher information matrix (CFIM) $\mathbf{C}=[C_{jk}]$ for computational-basis measurements has elements
\begin{equation}
C_{jk}
=
\sum_m
\frac{(\partial_jp_m) (\partial_kp_m)}{p_m},
\label{eq:CFIM}
\end{equation}
where $p_m$ denotes the probability of obtaining outcome $m$. The CFIM satisfies
$\mathbf{C}\preceq\mathbf{Q}$
for any measurement. In principle, $\mathbf{C}=\mathbf{Q}$ requires the symmetric logarithmic derivatives (SLDs) $L_j$ associated with each parameter to commute, $[L_j,L_k]=0$ ~\cite{Matsumoto_2002}. 

The variational circuits are optimized in two stages. We first optimize the state-preparation parameters $\bm{\theta}$ by minimizing $\Tr(\mathbf{Q}^{-1})$ at a fixed fiducial point, then, with $\bm{\theta}$ fixed, optimize the measurement parameters $\bm{\mu}$ by minimizing $\Tr(\mathbf{C}^{-1})$, the standard multiparameter Cram\'er-Rao bounds.

If the axion phase $\alpha$ were known, estimating the axion-photon coupling would reduce to a single-parameter problem in $\delta$, with the QFI and CFI as the natural figures of merit~\cite{rv43-54zq}. In practice, however, $\alpha$ is unknown and must be treated as an additional parameter. A straightforward strategy is therefore to estimate $(\delta,\alpha)$ jointly. This strategy, however, fails because the information about $\alpha$ vanishes in the weak-signal limit. Physically, as $\delta\rightarrow0$, the axion-induced rotation approaches the identity, making its rotation axis, and hence the phase $\alpha$, unresolvable. Consequently,
\begin{equation}
Q_{\alpha\alpha}(\delta,\alpha_0)
\simeq
c(\alpha_0)\,\delta^2,
\label{eq:Qalphacollapse}
\end{equation}
where $c(\alpha_0)\approx67$ is nearly independent of $\delta$, varying by less than $1\%$ over five orders of magnitude, and depends only weakly on the fiducial phase $\alpha_0$, varying by about $3\%$ across its full range; $Q_{\delta\delta}\approx67$ is comparable in size. Since realistic axion couplings correspond to $\delta\ll1$ [Eq.~\eqref{eq:delta2g}], the QFIM becomes nearly singular, yielding $\Tr(\mathbf{Q}^{-1})\approx3.5\times10^{8}$ regardless of the variational ansatz.

The singular QFIM does not imply a loss of physical information. Instead, it results from the choice of coordinate system. The variables $(\delta,\alpha)$ are simply the polar coordinates of
\(
(x,y)
=
(\delta\cos\alpha,\,
\delta\sin\alpha),
\)
which represent the same physical encoding. The corresponding Jacobian,
$
\big|
\frac{\partial(x,y)}
{\partial(\delta,\alpha)}
\big|
=\delta,
$
vanishes as $\delta\to0$. Thus, the polar coordinates become singular at the origin even though the physical unitary $U_{\rm DM}$ remains analytic at $(x,y)=(0,0)$. The collapse of $Q_{\alpha\alpha}$ in Eq.~\eqref{eq:Qalphacollapse} is therefore a coordinate singularity rather than a physical one.

Motivated by this observation, we reformulate the estimation problem in terms of the Cartesian variables $(x,y)$. Specifically, we replace the polar coordinates by
$\delta=\sqrt{x^2+y^2}$ and
$\alpha=\operatorname{atan2}(y,x)$,
and evaluate $\mathbf{Q}$ and $\mathbf{C}$ directly with respect to $(x,y)$. In this representation, these matrices remain well conditioned throughout the weak-signal regime: both eigenvalues of $\mathbf{Q}$ remain $\approx67$, comparable to $Q_{\delta\delta}$ above, and are essentially independent of the fiducial phase $\alpha_0$. As a result, ${\rm Tr}(\mathbf{Q}^{-1})\approx0.030$, about ten orders of magnitude smaller than in the polar parameterization at the same physical point, translating into a comparable, ten-order-of-magnitude reduction in the exposure required for a $5\sigma$ discovery of $\delta_0$.

In practice, we reconstruct $(\hat{x},\hat{y})$ from simulated click data using a two-dimensional Bayesian posterior and obtain the magnitude estimator
$\hat{\delta}=\sqrt{\hat{x}^2+\hat{y}^2}$
only as a final post-processing step. When $\delta$ becomes comparable to or smaller than the noise level in the $(x,y)$ plane, $\hat{\delta}$ exhibits a small bias. 
However, this is a standard finite-exposure effect and does not represent a fundamental loss of information or parameter degeneracy.

We optimize the variational parameters $\bm{\theta}$ and $\bm{\mu}$ by minimizing 
${\rm Tr}(\mathbf{Q}^{-1})$ and
${\rm Tr}(\mathbf{C}^{-1})$,
evaluated at the fiducial point
$(x_0,y_0)=(\delta_0,0)$.
Here,
\(
\delta_0\equiv\delta(g_{\rm KSVZ},m_a=1~\mu{\rm eV})
=6.499\times10^{-6},
\)
corresponds to the KSVZ benchmark interaction strength at
$m_a=1~\mu{\rm eV}$.
After this optimization, the trained protocol is applied to arbitrary dark-matter masses and phases. For each given $(m_a,\alpha)$, we reconstruct the Cartesian coordinates $(\hat{x},\hat{y})$ using Bayesian estimation and obtain the coupling estimator $\hat g_{a\gamma\gamma}$ through the recovered magnitude
$\hat{\delta}=\sqrt{\hat{x}^{2}+\hat{y}^{2}}$.

\begin{figure}[t]
\centering
\includegraphics[width=\columnwidth]{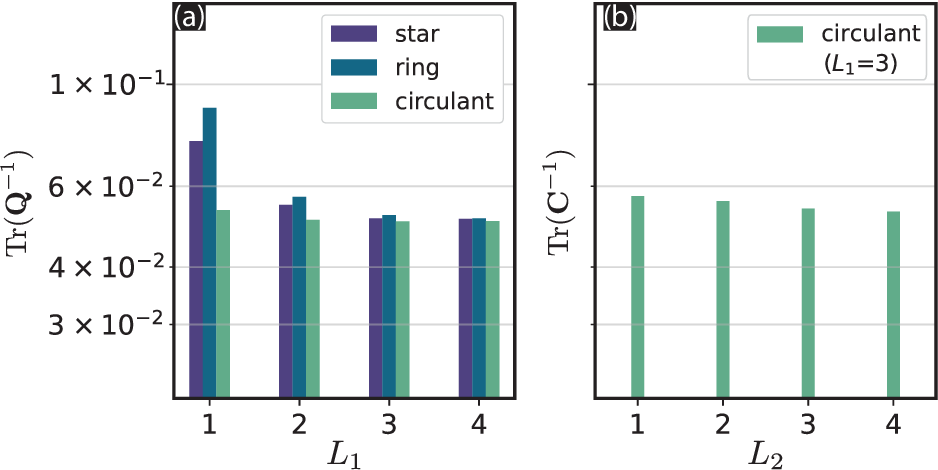}
\caption{\textbf{Cram\'er-Rao bounds as a function of circuit depth for $N=5$ at the fiducial point $(x_0,y_0)=(\delta_0,0)$.} 
(a) Quantum bound ${\rm Tr}(\mathbf{Q}^{-1})$ versus the state-preparation depth $L_1$ for the star, ring, and circulant topologies. 
(b) Classical bound ${\rm Tr}(\mathbf{C}^{-1})$ versus the measurement-circuit depth $L_2$ for the circulant topology, with $L_1=3$ fixed. 
Both bounds saturate at approximately three layers, leading to the choice $L_1=L_2=3$ in the subsequent analysis.
}
\label{fig:2}
\end{figure}

\inlinesection{Performance evaluation}
We first determine the required variational depth. Figure~\ref{fig:2}(a) shows the quantum bound ${\rm Tr}(\mathbf{Q}^{-1})$ as a function of the state-preparation depth $L_1$ for the three network topologies. The bound decreases rapidly from $L_1=1$ and reaches saturation around $L_1\sim3$, with the circulant network achieving the lowest value. We then fix the circulant topology with $L_1=3$ and evaluate the classical bound ${\rm Tr}(\mathbf{C}^{-1})$ as a function of the measurement-circuit depth $L_2$, as shown in Fig.~\ref{fig:2}(b). The classical bound similarly saturates at $L_2\sim3$. Therefore, we adopt $L_1=L_2=3$ for the circulant network in the following analysis.

\begin{figure}[t]
\centering
\includegraphics[width=\columnwidth]{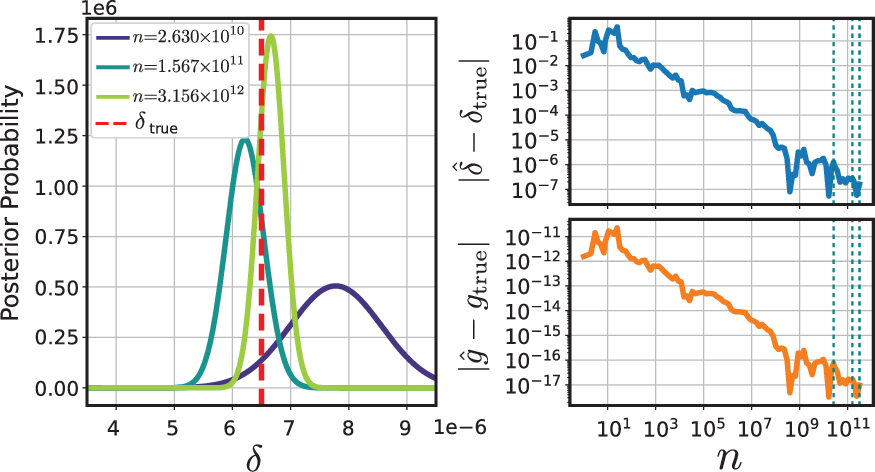}
\caption{
\textbf{A sequential Bayesian run up to a one-year exposure ($N_{\rm try}=3.156\times10^{11}$).}
(a) Marginal posterior distribution of $\delta$ at three checkpoints, one month, six months, and one year. The vertical dashed line indicates the injected value $\delta_{\rm true}$.
(b) Estimation error $\hat{\delta}-\delta_{\rm true}$ versus accumulated shots $n$, from $n=1$ to $N_{\rm try}$.
(c) Estimation error $\hat{g}_{a\gamma\gamma}-g_{\rm true}$ versus $n$.
The posterior progressively narrows as $n$ grows, and both estimation errors shrink in magnitude toward the injected values.
}
\label{fig:3}
\end{figure}

We fix the variational parameters $\bm{\theta}$ and $\bm{\mu}$ at their optimal values for the circulant network. Bayesian estimation is then performed for a signal with magnitude $\delta_{\rm true} = 6.499\times10^{-6}$ ($m_a=1~\mu{\rm eV}$) and a random phase $\alpha$. Measurement outcomes are accumulated over a total exposure corresponding to one year,
$N_{\rm try}=3.156\times10^{11}$,
and the running posterior is recorded after one month
($2.63\times10^{10}$ shots),
six months
($1.578\times10^{11}$ shots),
and one year.
The results are presented in Fig.~\ref{fig:3}.

As the number of accumulated measurements increases, the marginal posterior of $\delta$ [Fig.~\ref{fig:3}(a)] becomes progressively concentrated around the true value $\delta_{\rm true}$. Correspondingly, the estimation errors in $\hat{\delta}$ and $\hat{g}_{a\gamma\gamma}$ [Figs.~\ref{fig:3}(b) and \ref{fig:3}(c)] decrease by several orders of magnitude. After one year of integration, the residual error in $\hat{g}_{a\gamma\gamma}$ is approximately
$10^{-17}~{\rm GeV}^{-1}$,
which is many orders of magnitude smaller than the KSVZ benchmark coupling,
$g_{\rm KSVZ}\sim10^{-17}$--$10^{-15}~{\rm GeV}^{-1}$,
over the mass range considered.

\begin{figure}[t]
\centering
\includegraphics[width=\linewidth]{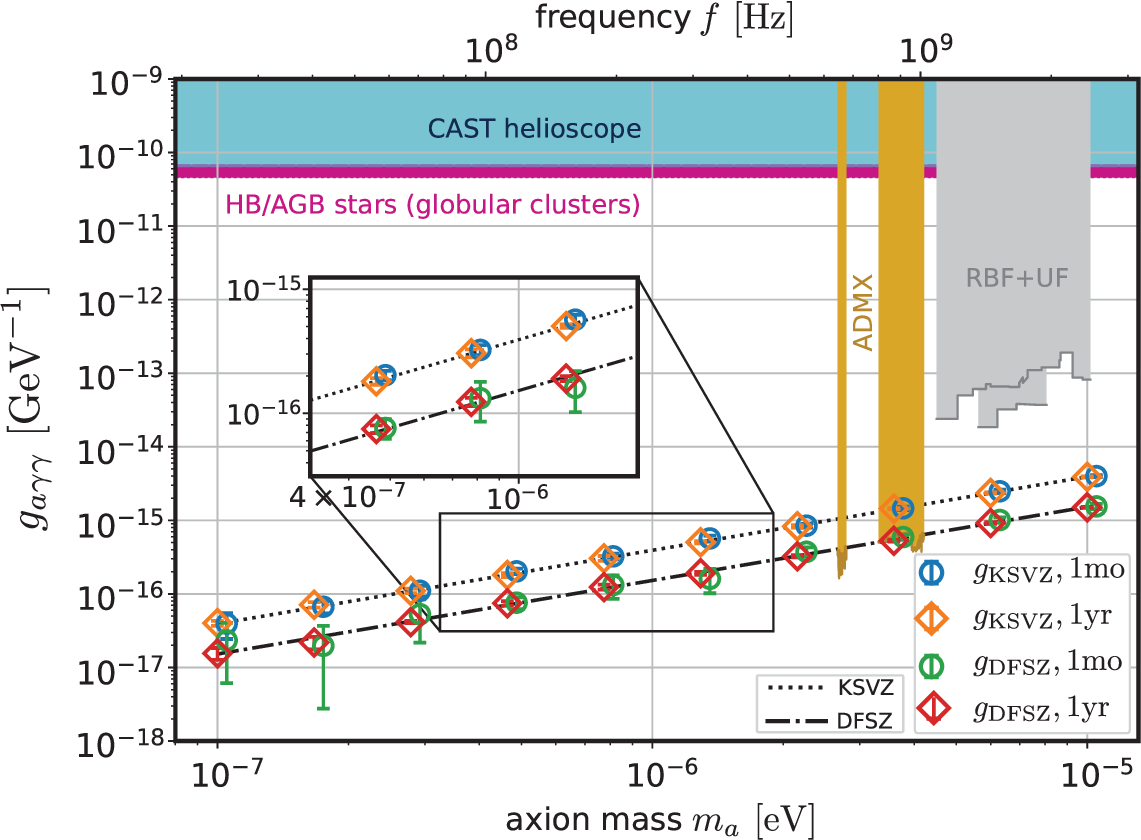}
\caption{
\textbf{Bayesian reconstruction of $g_{a\gamma\gamma}$ at ten target masses in the range
$m_a\in[0.1,10]~\mu{\rm eV}$.}
The reconstructed couplings $\hat g_{a\gamma\gamma}$ are shown for injected KSVZ and DFSZ benchmark signals under one-month and one-year exposures; symbols denote the mean over repeated Bayesian trials and error bars indicate one standard deviation. The dotted and dash-dotted black curves correspond to the KSVZ and DFSZ benchmark models, while the shaded regions and horizontal lines indicate existing haloscope and astrophysical constraints. The inset enlarges the region near $m_a\sim1~\mu{\rm eV}$, where the reconstructed KSVZ and DFSZ couplings are most clearly separated.}
\label{fig:4}
\end{figure}

Having validated the estimator at a representative mass point, we next assess its performance over the full target range without further optimization. We consider ten target masses spanning $m_a\in[0.1,10]~\mu$eV, taking either the KSVZ or DFSZ benchmark coupling, $g_{\rm true}=g_{\rm KSVZ}(m_a)$ or $g_{\rm DFSZ}(m_a)$ [Eq.~\eqref{eq:gagg}], as the injected signal. For each mass, we perform five independent Bayesian reconstructions of $\hat g_{a\gamma\gamma}$ under two realistic exposures, a one-month scan ($N_{\rm try}=2.630\times10^{10}$ shots per mass bin) and a one-year scan ($N_{\rm try}=3.156\times10^{11}$ shots per mass bin), and report their empirical mean and standard deviation. The recovered couplings are shown in Fig.~\ref{fig:4}, together with existing experimental and astrophysical constraints.

For both exposure times, the recovered couplings closely track the injected KSVZ and DFSZ benchmarks across the entire mass range. As expected, the one-year exposure yields smaller uncertainties, consistent with the $1/\sqrt{N_{\rm try}}$ scaling of statistical precision. These results demonstrate that the trained protocol generalizes across the full mass range without retraining and accurately reconstructs the injected couplings. We emphasize that Fig.~\ref{fig:4} validates the reconstruction performance of the estimator rather than presenting a projected discovery reach, which would instead be obtained directly from the Fisher information without assuming a specific injected signal. The reconstructed benchmark couplings fall well below current astrophysical limits, including the CAST~\cite{Anastassopoulos2017} and globular-cluster cooling~\cite{Dolan_2022} bounds, and span regions already probed by haloscope experiments such as ADMX~\cite{PhysRevLett.120.151301,PhysRevLett.127.261803} and the RBF/UF cavity searches~\cite{PhysRevLett.59.839,PhysRevD.42.1297}.

Finally, we investigate the performance of the protocol in the presence of realistic hardware noise. We consider two noise sources for current superconducting-qubit devices: energy relaxation and dephasing with $T_1=T_2=100~\mu$s during the interrogation time $\tau$, together with a $0.5\%$ depolarizing error on every entangling gate in both the state-preparation and measurement circuits. Under this noise model, we repeat the variational optimization by first minimizing ${\rm Tr}(\mathbf{Q}^{-1})$ over the state-preparation parameters $\bm{\theta}$ and circuit depth $L_1$, and then minimizing ${\rm Tr}(\mathbf{C}^{-1})$ over the measurement parameters $\bm{\mu}$ and depth $L_2$.

\begin{figure}[t]
\centering
\includegraphics[width=\linewidth]{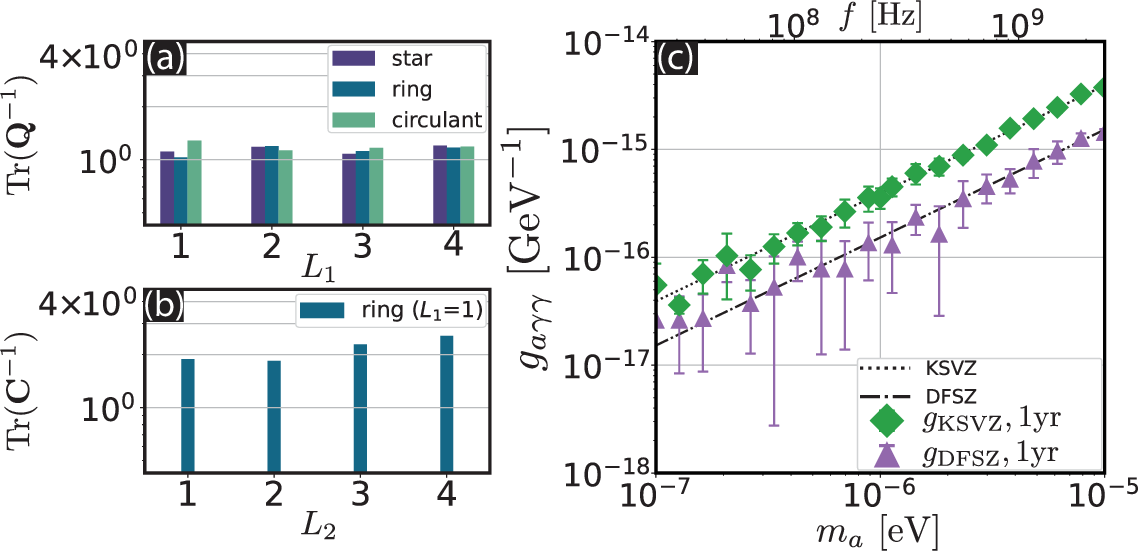}
\caption{
\textbf{Performance under a realistic noise model.} 
(a) Quantum bound ${\rm Tr}(\mathbf{Q}^{-1})$ versus the state-preparation depth $L_1$ for the star, ring, and circulant topologies.
(b) Classical bound ${\rm Tr}(\mathbf{C}^{-1})$ versus the measurement depth $L_2$ for the optimal topology selected in (a).
(c) Bayesian reconstruction of the axion-photon coupling for the optimized noisy circuit (ring topology with $L_1=1$ and $L_2=2$) using a one-year exposure. Symbols denote the reconstructed KSVZ and DFSZ benchmark couplings, while the dotted and dash-dotted curves indicate the corresponding theoretical models.
}
\label{fig:5}
\end{figure}

The results are summarized in Fig.~\ref{fig:5}. Panels (a) and (b) show that the ring topology becomes optimal under realistic noise because its reduced number of entangling gates results in lower accumulated gate error. The optimal noisy configuration is obtained for the ring topology with $L_1=1$ and $L_2=2$. Panel (c) presents the corresponding Bayesian reconstruction of the axion-photon coupling. Despite the increased statistical uncertainty, particularly at lower masses, the reconstructed couplings remain close to the injected KSVZ and DFSZ benchmarks. These results demonstrate that the variational optimization naturally adapts to realistic hardware noise by favoring shallower, less entangling circuits.

\inlinesection{Conclusions}
We have developed a variational quantum-sensor-network protocol for multiparameter detection of QCD axion dark matter using superconducting qubits in a cavity. By explicitly accounting for the unknown random dark-matter phase, we formulated axion detection as a two-parameter quantum estimation problem and identified a coordinate singularity that arises when the axion-induced qubit rotation is parameterized by its magnitude and phase. We showed that this singularity can be removed completely by reformulating the estimation problem in Cartesian coordinates while preserving the underlying physical encoding.

Within this framework, we optimized the quantum sensor network and demonstrated accurate Bayesian reconstruction of the axion-photon coupling across the target mass range. The optimized protocol successfully recovers the KSVZ and DFSZ benchmark couplings and remains robust against realistic superconducting-qubit decoherence and gate errors. Beyond its application to axion dark-matter detection, our work identifies the choice of estimation coordinates as an essential design element in multiparameter quantum sensing and establishes variational quantum sensor networks as a practical framework for quantum-enhanced sensing on near-term superconducting quantum platforms.

\inlinesection{Acknowledgments}
The code used to obtain the results in this Letter is publicly available at Ref.~\cite{Ho2026Code}. This work was supported by the Tohoku Initiative for Fostering Global Researchers for Interdisciplinary Sciences (TI-FRIS) under MEXT's Strategic Professional Development Program for Young Researchers.

\bibliographystyle{apsrev4-2}
\bibliography{refs}
\end{document}